\documentclass{raa}            
\usepackage{graphicx}
\usepackage{times}
\usepackage{natbib}
\bibpunct{(}{)}{;}{a}{}{,}
\usepackage{amssymb,amsmath}
\usepackage{float}
\usepackage{threeparttable}
\usepackage{booktabs}
\usepackage{adjustbox}
\usepackage{url}            
\usepackage[pagebackref]{hyperref}

\begin{document}

  \title{Warm Debris Disk Candidates around Nearby FGK Stars from LAMOST DR12
}

   \volnopage{Vol.0 (2000) No.0, 000--000}      
   \setcounter{page}{1}          

   \author{Xing Luo 
      \inst{1}
   \and Qiong Liu
      \inst{1}\thanks{Corresponding author, \url{https://orcid.org/0000-0001-6726-8907}}
}

   \institute{College of Physics, Guizhou University,
             Guiyang 550025, China; {\it qliu1@gzu.edu.cn}\\
\vs\no
   {\small Received 20xx month day; accepted 20xx month day}}

\abstract{Warm debris disks around main-sequence stars trace late-stage terrestrial planet formation. Motivated by the need for systematic searches of such systems, we identify debris disk candidates around FGK stars within 150 pc by combining a spectroscopically selected sample from LAMOST DR12 with \textit{Gaia} astrometry and multi-band infrared photometry. Infrared excesses are identified through SED fitting and validated using conservative, source-by-source checks. This approach yields a final sample of 12 debris disk candidates including ten new detections. Stellar age research indicate that most of the host stars are several billion years old. NEOWISE monitoring reveals no significant W1/W2 variability, consistent with a circumstellar origin of the infrared excess. while a search for co-moving companions using \textit{Gaia} DR3 reveals possible companions for only two candidates at very large projected separations ($\gtrsim 10^4$~au). Three candidates exhibit excess emission in both the W3 and W4 bands, allowing estimates of characteristic dust properties. This work establishes a small yet reliable sample of debris disk candidates anchored in homogeneous LAMOST spectroscopy, providing a foundation for future studies of debris disk evolution and stellar activity.
\keywords{stars: variables: general --- stars: solar-type --- (stars:) circumstellar matter --- infrared: planetary systems}
}

   \authorrunning{X. Luo \& Q. Liu }            
   \titlerunning{Verification of Infrared Excess in Debris Disk Candidates}  

   \maketitle

%
%
\section{Introduction}           
\label{sect:intro}
Debris disks are circumstellar dust structures generated by collisional cascades among planetesimals that failed to form planets, as well as by ongoing cometary activity \citep{2008ARA&A..46..339W}. They are commonly identified through infrared excess emission relative to stellar photospheres. As tracers of the late stages of planet formation and long-term dynamical evolution, debris disks around FGK stars are of particular interest for understanding the evolution of planetary systems, including the Solar System \citep{2018ARA&A..56..541H}.

Large photometric surveys such as WISE and Gaia have enabled statistical studies of infrared excess and debris disk occurrence rates around nearby stars \citep{2005ApJ...620.1010R,2014MNRAS.444.3164K,2024AJ....167..275M}. However, photometrically selected samples are susceptible to contamination (e.g., background sources and unresolved binaries) and offer limited constraints on stellar physical and activity properties, which are crucial for detailed investigations of debris disk evolution.

Large spectroscopic surveys provide a complementary approach by offering homogeneous and reliably classified stellar samples, as well as access to stellar physical and activity diagnostics \citep{2008ApJ...674.1086T, 2014ApJS..211...25C, 2017ApJ...848...41L}. The Large Sky Area Multi-Object Fiber Spectroscopic Telescope (LAMOST) has obtained millions of low- and medium-resolution spectra for Galactic stars, forming a comprehensive database suitable for both statistical and targeted studies \citep{2012RAA....12.1197C, 2012RAA....12..723Z, 2015RAA....15.1095L, 2020ApJ...891...23W}. LAMOST spectroscopy provides essential constraints on stellar physical properties and activity, while its large and homogeneous stellar catalog supports studies of Galactic archaeology, stellar ages \citep{2023A&A...673A.155Q}, and stellar magnetic activity \citep{2022ApJS..261...26S}, with implications for exoplanet habitability.

We use the LAMOST DR12 \emph{LAMOST LRS Stellar Parameter Catalogue of A, F, G, and K Stars}\footnote{\url{https://www.lamost.org/dr12/v1.0/catalogue}}. as the parent sample and select FGK stars based on color criteria. Infrared excesses are identified via spectral energy distribution (SED) fitting and confirmed through conservative, source-by-source checks. Stellar parameters are taken from the LAMOST, with stellar ages compiled from \cite{2023A&A...673A.155Q}. This work aims to establish a homogeneous sample of debris disk candidates with LAMOST spectroscopic observations, providing a robust basis for future studies of debris disk evolution and its dependence on stellar properties.

The paper is organized as follows. Section~\ref{sect:Sample} describes the sample construction and data selection. Section~\ref{sect:SED Fit} details the SED fitting procedure and infrared excess identification. Section~\ref{sect:Origins of Infrared Excess} presents a multi-step validation of the infrared excess to assess its circumstellar origin. The results are presented and discussed in Section~\ref{sect:Results and Discussion}, and the summarize work in Section~\ref{sect:Summary}.


\section{Sample Selection and Data Filtering}
\label{sect:Sample}

LAMOST is a north–south reflecting Schmidt telescope with a 5$^\circ$ field of view \citep{2012RAA....12..735D,2015RAA....15.1095L}. Its deformable Schmidt plate adapts to target declination during tracking, providing an effective aperture of 3.6–4.9 m \citep{1996ApOpt..35.5155W,2012RAA....12.1197C}. With 4000 fibers across its wide field, LAMOST can obtain parallel spectra down to 20.5 mag in 1.5 hr exposures.  
The initial sample is drawn from the LAMOST DR12 \emph{LAMOST LRS Stellar Parameter Catalogue of A, F, G, and K Stars}, comprising $\sim$8.37 million low-resolution ($R\sim1800$) stellar spectra with spectroscopic parameters (spectral type, $T_{\rm eff}$, $\log g$, $\mathrm{[Fe/H]}$, and RV) and cross-matched Gaia DR3 source IDs and $G$-band photometry. This dataset is hereafter referred to as \textbf{Sample A}.

Since the catalog includes multiple low-resolution observations of the same star, we first remove duplicate entries based on the LAMOST-provided designation ID (uid). Considering the high stellar density and potential contamination near the Galactic plane, stars with Galactic latitude $|b|<10^\circ$ are further excluded. In addition, stars classified as type A are removed, reducing the sample size to about 4.57 million sources.

As LAMOST is a spectroscopic survey and does not deliver astrometric parameters (e.g., parallax and proper motion), these 4.57 million sources are cross-matched with the Gaia DR3 catalog using the Gaia source ID provided by LAMOST (\texttt{gaia\_source\_id}), thereby obtaining parallaxes and other astrometric parameters. To ensure reliable distance estimates, only stars satisfying $\pi/\sigma_\pi>5$ are retained. Distances are estimated from the parallax, and stars within $d<150$\,pc are selected. After this selection, the sample is reduced to about 23,000 sources.

Using Gaia DR3 proper motion data from 2016, the positions of these stars were propagated to the 2010 epoch and cross-identified with the AllWISE catalog within a 1\arcsec\ radius using TOPCAT. Photometric data from both AllWISE and 2MASS were then retrieved. To select main-sequence stars, we applied photometric criteria to exclude evolved contaminants. Following \citet{2021ApJ...910...27M}, giant stars were removed using
\begin{equation}
M_J < 5\,(J-K_s),
\end{equation}
where $J$ and $K_s$ are the $J$ and $K_s$ band magnitudes of 2MASS, and $M_J$ is the absolute magnitude of 2MASS $J$ band.

White dwarfs were excluded according to \citet{2016ApJS..225...15C} using
\begin{equation}
M_V \ge 2.5\,(V_T-W2)+1.8,
\end{equation}
where $V_T$ is the Tycho $V_T$ magnitude, $W2$ is the AllWISE $W2$ band magnitude, and $M_V$ is the absolute magnitude of Tycho $V_T$ band.

Since not all targets have Tycho photometry, we estimated $V_T$ using the empirical Gaia DR3 photometric transformation relation\footnote{\url{https://gea.esac.esa.int/archive/documentation/GEDR3/Data_processing/chap_cu5pho/cu5pho_sec_photSystem/cu5pho_ssec_photRelations.html}}:
\begin{equation}
V_T = G + 0.01077 + 0.0682\,C + 0.2387\,C^2 - 0.02342\,C^3,
\end{equation}
where $G$ is the Gaia broad-band magnitude, $C = G_{\rm BP}-G_{\rm RP}$ is the Gaia color index, and $G_{\rm BP}$ and $G_{\rm RP}$ are the Gaia blue and red photometric magnitudes, respectively.

\begin{figure}
    \centering
    \includegraphics[width=0.5\linewidth]{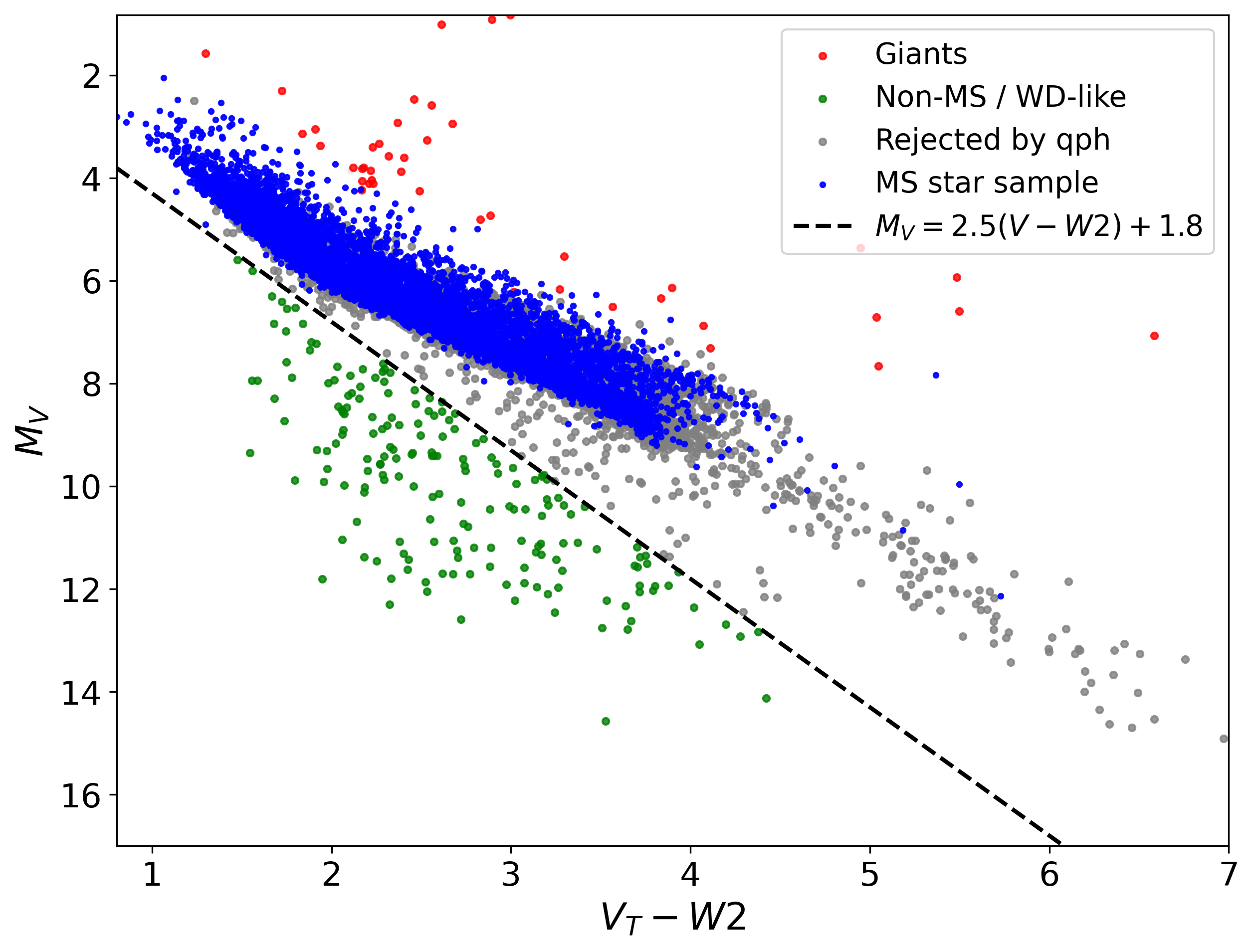}
    \caption{Distribution of the $\sim$23\,000 sources after a sequential selection process, with colors indicating excluded objects and the dashed line denoting the white dwarf removal criterion.
}
    \label{fig:MV_VS_J-W2}
\end{figure}

To further improve the reliability of infrared measurements, we perform quality control on AllWISE photometry by requiring \texttt{ccf=0000}, \texttt{ex=0}, and considering the \texttt{qph} quality flag, retaining only sources with the first three characters as ``AAA'' and the fourth as A, B, or C. After these selections, a total of 7,760 candidate sources are obtained, hereafter defined as \textbf{Sample B}, which serves as the working sample for subsequent debris disk analysis.Figure~\ref{fig:MV_VS_J-W2} shows the color--absolute magnitude diagram ($M_V$ vs. $V_T-W2$) of the filtered sample, resembling a Hertzsprung--Russell diagram.

\section{Verification of Infrared Excess}
\label{sect:SED Fit}
To determine whether a star exhibits infrared (IR) excess, we predict the stellar photospheric flux from the optical to infrared bands via SED fitting and compare it with observed fluxes. In this study, we employ the VO SED Analyzer (VOSA) developed and maintained by the Instituto de Astrof\'isica de Canarias (CAB/LAEFF), which operates within the Virtual Observatory (VO) framework~\citep{2008A&A...492..277B}. VOSA supports large-sample SED fitting and IR excess identification, and allows the selection of atmospheric models tailored to the stellar parameters, ensuring reliable fitting results.

The initial sample consists of FGK stars with multi-band photometry compiled from Gaia DR3, 2MASS, and AllWISE catalogues. Refer to the stellar parameters Teff and logg provided by LAMOST, during the fitting procedure, interstellar extinction is automatically determined by VOSA. Theoretical models from Kurucz (2003, $\alpha=0.0$) are adopted \citep{2003IAUS..210P.A20C}, with surface gravity $\log g$ in the range $4-5$ dex and metallicity [Fe/H] fixed between $-1$ and $0.5$ dex. Based on the SED fitting results, we extract the predicted stellar fluxes in the W3 and W4 bands ($F_{12,\rm phot}$, $F_{22,\rm phot}$) and compare them with the observed fluxes ($F_{12,\rm obs}$, $F_{22,\rm obs}$) and associated observational uncertainties ($\sigma_{12,\rm tot}$, $\sigma_{22,\rm tot}$) to define the excess flux:
\begin{equation}
F_{\rm excess} = F_{\rm obs} - F_{\rm phot}.
\end{equation}
The excess is considered significant if
\[
\chi = \frac{F_{\rm excess}}{\sigma_{\rm tot}} \ge 3,
\quad
\sigma_{\rm tot} = \sqrt{\sigma_{\rm obs}^{2} + \sigma_{\rm cal}^{2}} .
\]
Here, $\sigma_{\rm obs}$ denotes the observational uncertainty in W3 or W4, while $\sigma_{\rm cal}$ represents the calibration uncertainty arising from the telescope's absolute calibration~\citep{2011ApJ...735..112J}. The absolute calibration uncertainties of the WISE telescope are $\sim 4.5\%$ for W3 ~\citep{2014MNRAS.437..391C}and $\sim 5.7\%$ for W4~\citep{2011ApJ...735..112J}.
\begin{equation}
\sigma_{\rm cal} = F_{12,\rm obs} \times 0.045 \quad {\rm (W3)}, \quad 
\sigma_{\rm cal} = F_{22,\rm obs} \times 0.057 \quad {\rm (W4)}.
\end{equation}

After this validation and filtering procedure, 51 sources are found to exhibit significant IR excess in either the W3 or W4 band. These 51 objects are defined as the IR excess sample, hereafter referred to as Sample~C. Further systematic checks will be performed on Sample~C to select those sources whose IR excess is attributable to circumstellar debris discs.

\section{Assessment of the Origins of Infrared Excess}
\label{sect:Origins of Infrared Excess}
Since infrared excess can arise from various sources, we focus on identifying  FGK stars with debris discs. This section outlines the verification procedure, including image inspection, binary search, positional offset analysis, and the use of a matching Figure of Merit (FoM) to minimize misidentifications and contamination, ensuring reliable candidates.

\subsection{Visual Inspection}
\label{sect:Visual Inspection}
Although modern data pipelines use automatic flags (e.g., contamination flags  \texttt{'ccf'} and extended source flags  \texttt{'ex'}) to improve efficiency, these mechanisms have limitations. Artifacts, blended sources, and complex backgrounds may not be properly identified, potentially affecting analysis reliability. Therefore, manual visual inspection is an essential step in the data screening process.

~\citet{2024MNRAS.531..695S} summarized three main types of residual contaminants that are often missed in automatic selection:
\begin{itemize}
   \item Blended Sources: Contamination from target overlap with nearby sources in WISE images, especially in W3 and W4 bands. High-resolution optical images can help identify these overlaps. If a contaminant lacks optical emission and the IR source is offset without an optical counterpart, it is classified as a blend and excluded.
\item Nebular Features: Diffuse, complex structures in W3 and W4 images where the candidate’s IR core is unclear. Large-scale images ($\sim600$~arcsec) reveal extended nebular structures misidentified as point sources.
\item Irregular Sources: WISE W3 and W4 sources deviating from point-like morphology. Despite \texttt{ext\_flag} = 0, these sources show no nebular structures on large scales. Their irregularity may be due to faint nebular remnants, high noise, or blends, though the cause remains unclear.
\end{itemize}
Combining Pan-STARRS optical images with AllWISE IR images, we performed a joint visual inspection and identified 30 contaminated or morphologically anomalous sources: 11 sources show evidence of potential blending, with nearby infrared sources located within 12~arcsec that may affect the measured fluxes. 14 sources exhibiting nebular-like structures in W3 or W4, and 5 irregular sources in W4. The visual inspection of the excluded 30 sources is shown in the appendix \ref{sec:Sources Excluded Based on Visual Inspection}. 21 IR excess candidates remained from Sample~C.

\subsection{Close Binaries}

Close binaries, particularly those with significant luminosity or temperature differences between components, can produce apparent IR excess relative to single-star models. To assess the 21 candidate sources, we follow ~\citet{2019RAA....19...64Q} based on LAMOST survey results: if the radial velocity of a target differs by more than 10~km~s$^{-1}$ across different epochs, it is considered a close binary candidate. 

In addition to RV variability, we examine the Gaia DR3 Renormalised Unit Weight Error (RUWE), which quantifies the quality of the single\,star astrometric solution~\citep{2022MNRAS.517.2925C}. Values close to unity indicate a good fit, whereas RUWE $>1.4$ suggests unresolved binarity or orbital motion.

Among the remaining 21 sources, Gaia DR3~3088740475546133888 has a RUWE of 2.93691, significantly above the ideal single-star value, indicating large residuals in the Gaia single-star astrometric model fit. Furthermore, its radial velocities measured by Gaia and LAMOST differ markedly. LAMOST reports three radial-velocity measurements at MJD~56315 and 56667, yielding values of $73.62 \pm 3.77$~km~s$^{-1}$ and $86.51 \pm 5.91$~km~s$^{-1}$, respectively. In contrast, Gaia DR3 reports a radial velocity of $88.40 \pm 0.17$~km~s$^{-1}$ at the Gaia ref\_epoch of 2016. The significant RV variability observed in the LAMOST measurements, together with the discrepancy between Gaia and LAMOST radial velocities and the elevated RUWE value, strongly suggests that the source is an unresolved or close multiple system. To avoid potential bias, this source is excluded from subsequent debris disc analysis.

\subsection{Positional Offset Analysis}
To rule out coincidental alignments or background contamination, we follow \citet{2016ApJS..225...15C} and retrieve $3'\times3'$ AllWISE images in four bands for the remaining 20 IR excess sources  from IRSA. Source positions in each band are extracted using \texttt{DAOStarFinder} with parameters: \texttt{fwhm=1}, \texttt{sigma=0.1}, \texttt{threshold=3.1}. Offsets are calculated relative to W2 positions, with thresholds of 6.7~arcsec for W3 and 12~arcsec for W4. Result shows 2 sources exceed the W3 threshold, and 6 sources exceed the W4 threshold. After this step, 12 sources remain, whose positional offsets are summarized in Table~\ref{tab:pos_offset}.

\begin{table}[htbp]
\centering
\caption{The positional offsets of the 12 debris disk candidates in the W3 and W4 bands relative to W2.}
\begin{tabular}{l l c c}
\toprule
SIMBAD & AllWISE & W3,W2($^{\prime\prime}$) & W4,W2($^{\prime\prime}$) \\
       &         & RA,DEC                  & RA,DEC                  \\[-0.8ex]
\midrule
TYC 3422-922-1 & J083717.38+512013.7 & -1.7056 ,  2.4553 &  1.5067 ,  2.3286 \\
---            & J022959.14+362405.6 & -1.2865 ,  2.8960 &  4.3958 ,  2.5339 \\
HD 22680       & J033941.18+231726.7 & -0.0300 ,  0.1969 & -0.3028 ,  1.8534 \\
TYC 2799-987-1 & J010859.11+385050.9 &  0.7855 ,  1.4481 & -1.5380 ,  2.0214 \\
StKM 1-140     & J011938.78+191543.8 &  0.3668 , -1.3617 &  2.1934 ,  1.2580 \\
HD 283427      & J040814.09+243809.5 & -2.5270 , -0.3412 & -1.3932 ,  0.8295 \\
BD+46 1412     & J083128.30+454352.2 &  0.5818 ,  0.1690 &  0.5013 , -1.2551 \\
TYC 814-2432-1 & J085323.86+101542.7 &  1.2098 , -1.0164 &  1.4166 , -1.4167 \\
TYC 3885-281-1 & J163811.37+575551.4 &  0.7409 , -2.0042 & -7.0973 , -8.2177 \\
TYC 3845-1057-1 & J124757.84+565529.5 & -0.7888 , -1.0597 & -8.6604 , -3.9462 \\
TYC 1450-685-1 & J130540.51+163827.2 &  1.4920 , -0.4109 & 11.0223 , -4.4145 \\
2MASS J15353953+1922552 & J153539.53+192253.8 & -0.7352 , -0.7151 &  7.8043 ,  1.8065 \\
\bottomrule
\end{tabular}
\label{tab:pos_offset}
\end{table}

\subsection{Figure of Merit}

\begin{figure}[H]
    \centering
    \includegraphics[width=0.7\linewidth]{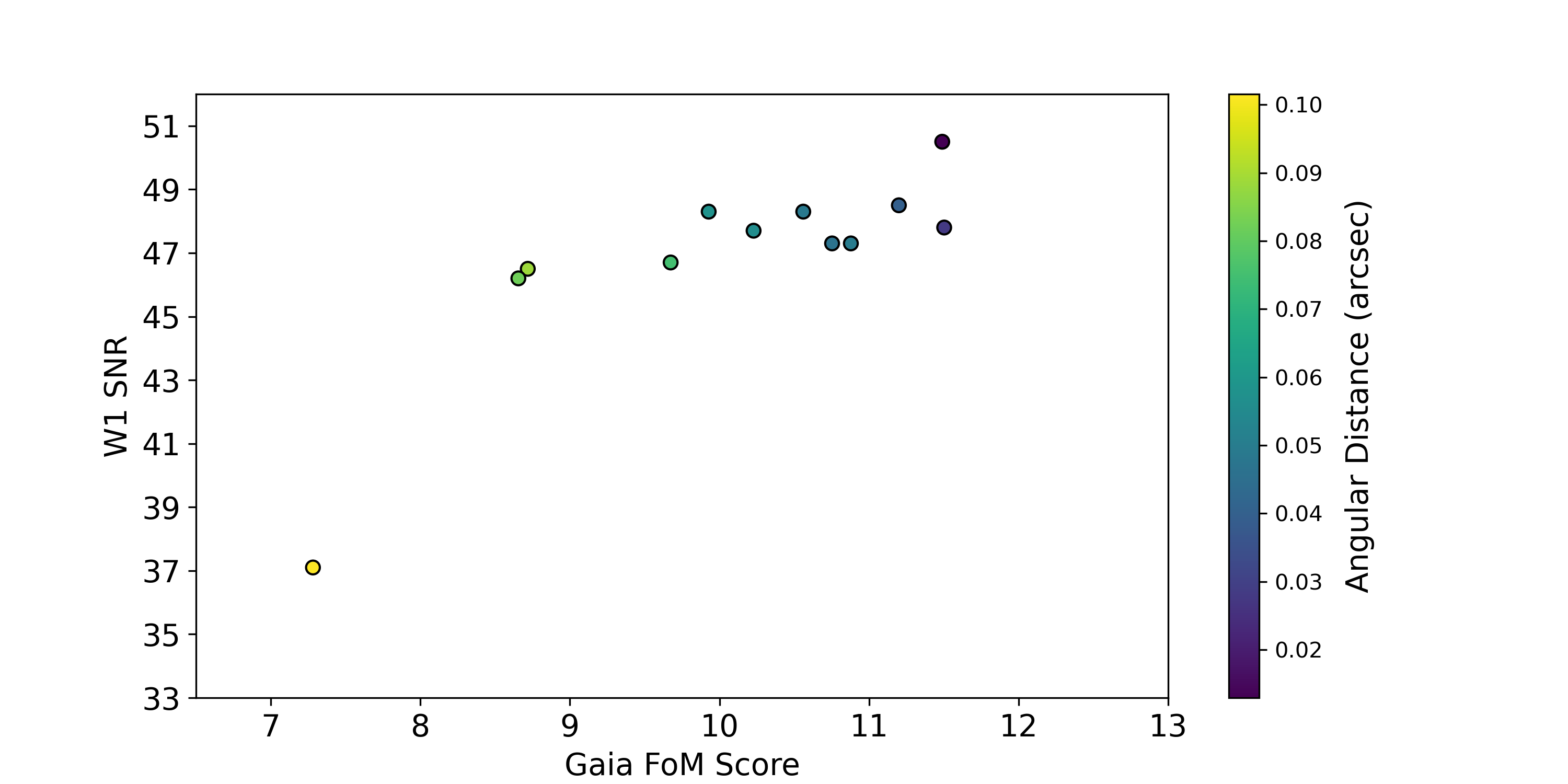}
    \caption{The relationship between the Gaia FoM and the signal-to-noise ratio (S/N) in the AllWISE W1 band for the 12 debris-disk candidates, with the color scale showing the positional offset between the Gaia and AllWISE positions.}
    \label{fig:FoM}
\end{figure}

The Figure of Merit (FoM) is a dimensionless score provided by the Gaia cross-matching algorithm when associating sources with other catalogues, quantifying the reliability of matches. Higher FoM values indicate more reliable matches. According to ~\citet{2020ApJ...891...97D}, if a target exhibits a low FoM value while having a high signal-to-noise ratio (SNR), its IR flux is likely contaminated by nearby sources rather than originating from the target itself. When comparing with WISE data, sources with SNR $>10$ and FoM $<4$ are recommended to be excluded. After screening, all 12 targets lie outside this exclusion zone, satisfying reliability criteria. For convenience, we define this set as Sample~D. Figure~\ref{fig:FoM} shows the distribution of SNR versus FoM in the W1 band.

\section{Results and Discussion}
\label{sect:Results and Discussion}
\subsection{Results}
\begin{figure}[H]
    \centering
    \includegraphics[width=0.8\linewidth]{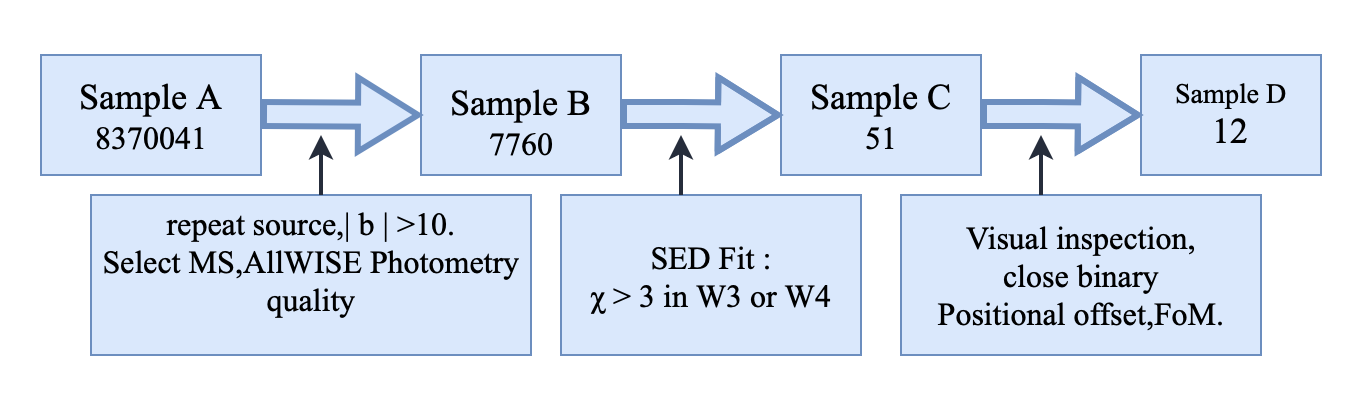}
    \caption{Sample Selection and Screening Process}
    \label{fig:Flow_chart}
\end{figure}

\begin{figure}[H]
    \centering
    \includegraphics[width=1\linewidth]{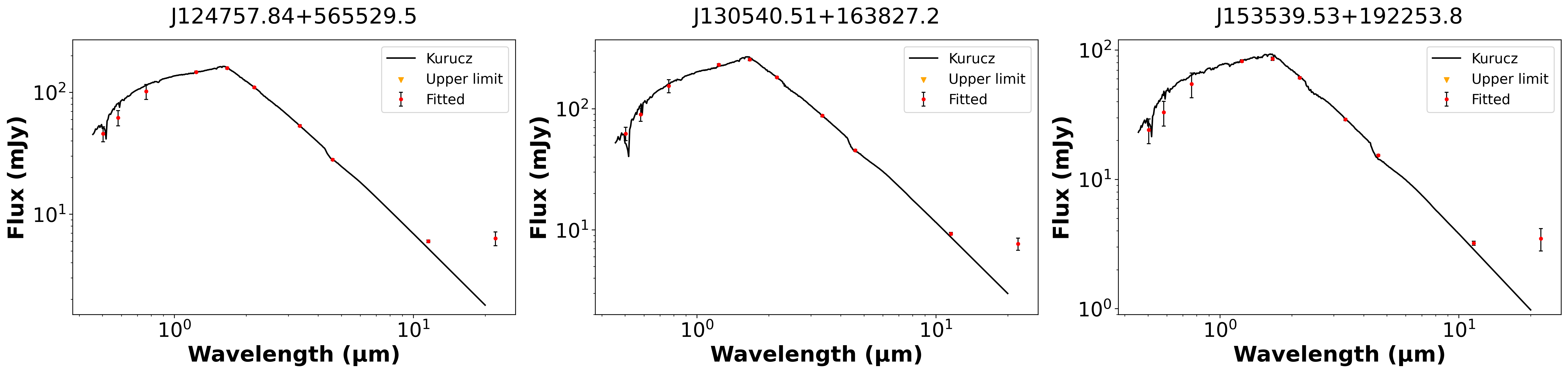}
    \caption{For the fitting, all targets use photometric data from Gaia DR3, 2MASS, and AllWISE, except for J011938.78+191543.8, for which additional optical data from SDSS were included.}
    \label{fig:SED}
\end{figure}

After verification of the infrared excesses and subsequent quality checks, we finally confirm a robust sample of debris disk candidates associated with FGK stars. referred to as Sample~D. The stellar parameters derived from the LAMOST, together with stellar ages compiled from \citep{2023A&A...673A.155Q}, are summarized in Table~\ref{tab:stellar_param}. Among these sources, three exhibit infrared excesses in both the W3 and W4 bands (J083717.38+512013.7, J022959.14+362405.6, J033941.18+231726.7), while the remaining nine show excess only in the W4 band. Figure~\ref{fig:Flow_chart} illustrates the sample selection and screening procedure adopted in this work.
Figure~\ref{fig:SED} shows the SEDs of the 12 confirmed debris disc candidates. For J040814.09+243809.5, we retrieved optical fluxes from Tycho-2, SDSS, and Gaia DR3, and found that all of them are systematically lower than the model fluxes. In order to achieve a better fit, we therefore replaced the Kurucz model with the BT-Settl model.A comparison of the stellar parameters derived from LAMOST and SED fitting is summarized in Table ~\ref{tab:stellar_param}.

We note that the surface gravities derived from LAMOST spectroscopy and SED fitting differ by up to $\sim 0.75$ dex for some sources. The LAMOST parameters are obtained with the LAMOST Stellar Parameter Pipeline (LASP) via full-spectrum fitting using the Université de Lyon Spectroscopic analysis Software (ULySS) \citep{2011RAA....11..924W,2015RAA....15.1095L}, where $\log g$ is constrained by pressure-sensitive features such as the Balmer line wings and the Mg~I~b triplet, with typical uncertainties of $\sim 0.1$--0.3 dex.
In contrast, SED fitting relies on broadband photometry and is mainly sensitive to $T_{\rm eff}$, providing only weak constraints on $\log g$ \citep{2011MNRAS.411..435B}.Therefore, offsets between spectroscopic and SED-derived surface gravities are not unexpected.
All targets remain within the main-sequence regime ($\log g \sim 4.0\!-\!4.75$), and these differences do not affect their evolutionary classification or the conclusions of this work.

\begin{figure}[H]
    \centering
    \includegraphics[width=1\linewidth]{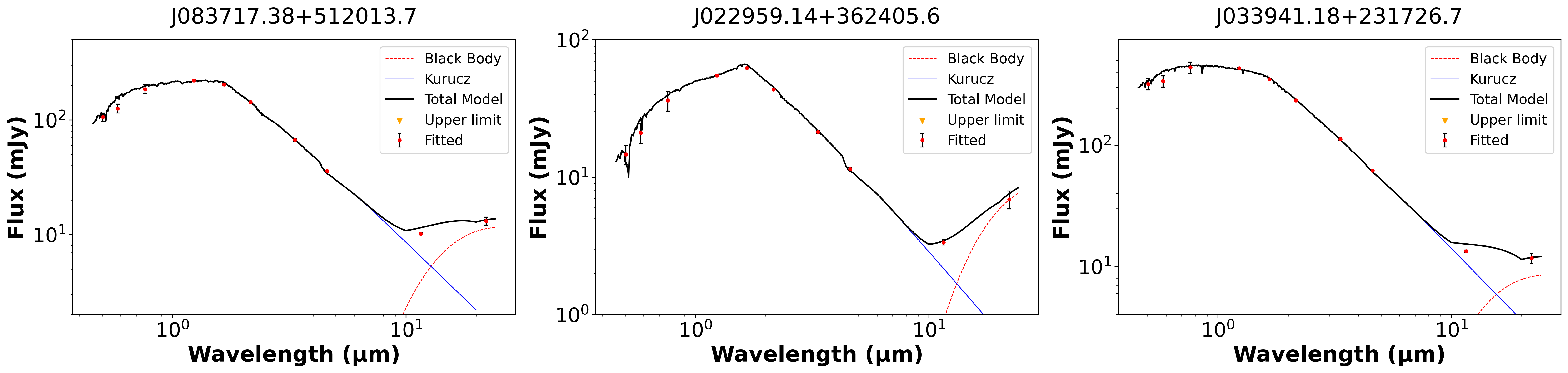}
    \label{fig:SED1}
\end{figure}
\begin{figure}[H]
    \centering
    \includegraphics[width=1\linewidth]{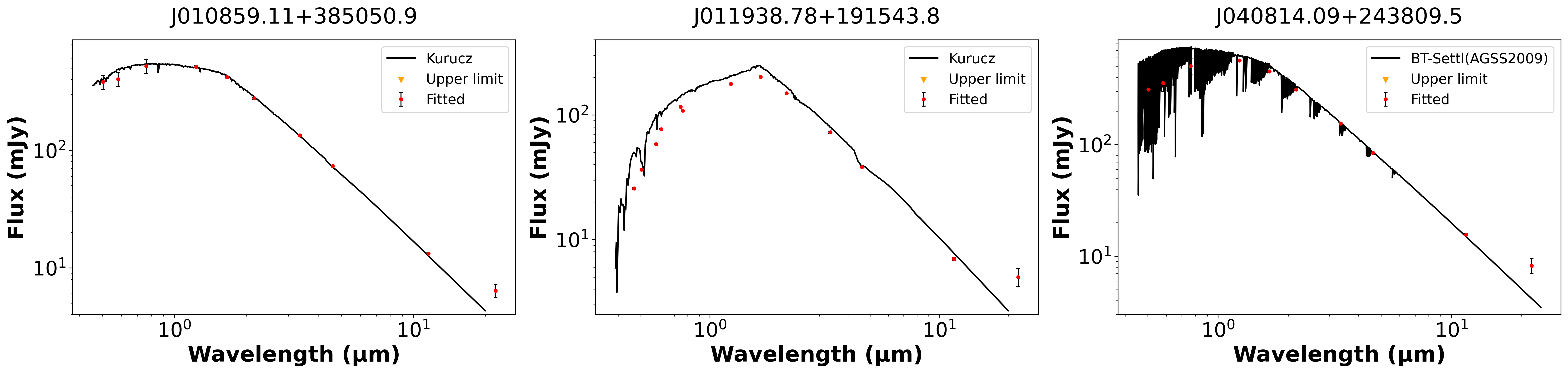}
    \label{fig:SED2}
\end{figure}
\begin{figure}[H]
    \centering
    \includegraphics[width=1\linewidth]{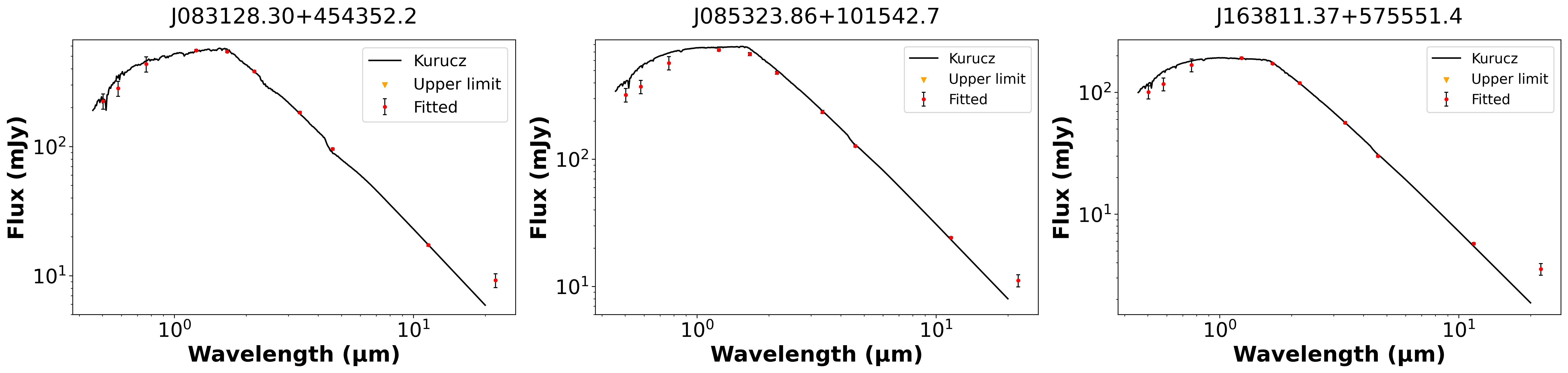}
    \label{fig:SED3}
\end{figure}

\begin{table*}[htbp]
\centering
\caption{Stellar parameters of the 12 debris disk candidates.}
\label{tab:stellar_param}

\setlength{\tabcolsep}{3.5pt}
\renewcommand{\arraystretch}{0.95}

\begin{threeparttable}
\footnotesize
\begin{adjustbox}{width=\textwidth,center}
\begin{tabular}{l c c c c c c c c c c c c}
\toprule
AllWISE
& $T_{\rm eff}$ & $T_{\rm eff}^{*}$ & SpT
& $\log g$ & $\log g^{*}$ & [Fe/H] & RV
& $G$ & RUWE & D & Age & Age$_{\rm 1}$ \\
& (K) & (K) &
& (dex) & (dex) & (dex) & (km\,s$^{-1}$)
& (mag) &  & (pc) & (Gyr) & (Gyr) \\
\midrule
J083717.38 & 5399 & 5250 & G8 & 4.76 & 4.0 &  0.166 &   5.20 & 10.878 & 0.901 & 122.2 &  8.55 &  5.10 \\
J022959.14 & 4483 & 4500 & K5 & 4.69 & 5.0 & -0.495 & -30.77 & 12.821 & 0.969 & 144.5 & \dots & \dots \\
J033941.18 & 6253 & 6000 & F7 & 4.38 & 4.0 &  0.063 &   5.08 &  9.811 & 0.847 & 138.1 &  4.25 & \dots \\
J010859.11 & 6181 & 6000 & F7 & 4.38 & 4.0 &  0.056 &   1.20 &  9.620 & 0.862 & 123.4 &  4.20 &  4.04 \\
J011938.78 & 4111 & 4500 & K5 & 4.49 & 5.0 & -0.206 &   3.92 & 11.716 & 1.378 &  81.6 &  3.04 & 12.70 \\
J040814.09 & 6323 & 6300 & F7 & 4.23 & 4.0 &  0.251 &  36.51 &  9.751 & 1.032 & 142.9 &  4.76 &  0.63 \\
J083128.30 & 5088 & 5000 & G9 & 4.75 & 4.0 &  0.088 &   6.89 & 10.003 & 1.256 &  63.9 & 12.84 &  0.12 \\
J085323.86 & 5421 & 5500 & F9 & 4.63 & 4.0 &  0.081 &   8.37 &  9.701 & 0.914 &  70.7 &  0.35 & \dots \\
J163811.37 & 5416 & 5500 & G5 & 4.40 & 5.0 &  0.213 & -66.33 & 10.959 & 0.938 & 146.0 &  8.60 &  3.90 \\
J124757.84 & 4494 & 4750 & K3 & 4.56 & 5.0 & -0.542 &  34.22 & 11.651 & 1.007 & 104.3 &  3.33 & 13.23 \\
J130540.51 & 4412 & 4500 & K5 & 4.62 & 5.0 & -0.112 &  13.83 & 11.246 & 1.389 &  79.1 &  4.09 &  7.94 \\
J153539.53 & 4615 & 4750 & K4 & 4.63 & 4.0 & -0.321 & -13.92 & 12.334 & 1.133 & 145.0 & 10.59 & \dots \\
\bottomrule
\end{tabular}
\end{adjustbox}

\begin{tablenotes}
\footnotesize
\item \textit{Notes.}
The AllWISE ID is abbreviated. For details, refer to the Table \ref{tab:pos_offset}, $T_{\rm eff}^{*}$ and $\log g^{*}$ are the adopted stellar parameters used for the SED fitting.
$T_{\rm eff}$, $\log g$, [Fe/H], and RV are taken from the LAMOST. The $G$-band magnitude, RUWE, distance, and stellar age are obtained from \textit{Gaia}. Age$_{\rm 1}$ denotes the stellar age derived by \citet{2023A&A...673A.155Q} based on the LAMOST DR7 catalog.
\end{tablenotes}
\end{threeparttable}
\end{table*}

\subsection{Discussion}
\subsubsection{Comparison with Previous Results}
For the final sample of 12 confirmed debris disk candidates, we conducted a systematic literature search using the SIMBAD database. We found that J033941.18+231726.7 and J130540.51+163827.2 were previously identified as debris disk systems by \citet{2010ApJ...712.1421S} and \citet{2016ApJS..225...15C}, respectively. The remaining 10 candidates have not been reported as debris disks in the literature and may represent previously unrecognized systems.

While both this work and \citet{2024AJ....167..275M} utilize \textit{Gaia} and \textit{WISE} data, no overlap is found between the two samples. The study by \citet{2024AJ....167..275M} relies on a large, photometrically selected sample intended for statistical analysis of debris disk occurrence, focusing on broad population trends. whereas our study adopts a conservative, source-by-source identification strategy based on a spectroscopically defined stellar sample from LAMOST. Consequently, the two studies probe complementary regions of the \textit{Gaia}–\textit{WISE} parameter space.

Comparisons with previous photometric searches \citep{2016ApJS..225...15C,2014ApJS..212...10P,2014MNRAS.437..391C} show little or no overlap with our 12 candidates. These earlier studies relied primarily on color-based selection to construct large or magnitude-limited samples, whereas our analysis emphasizes spectroscopically derived stellar properties and a more conservative, source-by-source validation process, including SED fitting and \textit{Gaia}-based reliability checks. Thus, our sample complements those identified in earlier photometric surveys.

Similarly, \citet{2017ApJ...837...15D} focused on \textit{WISE} infrared excess in a sample of solar analogs and twins, with an emphasis on population-level statistics. In contrast, our work considers a broader sample of FGK main-sequence stars and applies a conservative identification strategy using a large spectroscopic database, providing a complementary perspective on debris disks around Sun-like stars.

\subsubsection{Infrared Variability in W1 and W2}

To further assess debris disk candidates, temporal infrared flux variability is used as a complementary diagnostic to infrared excess measurements. Long-term WISE monitoring in the 3.4~$\mu$m (W1) and 4.6~$\mu$m (W2) bands enables an investigation of variability associated with stars. These bands are dominated by stellar photospheric emission, with possible contributions from warm inner dust. As a result, infrared variability may arise from stellar luminosity or activity variations, as well as from changes in circumstellar dust properties, such as dust production or removal, collisional events, or short-term occultations.

Following \citet{2021ApJ...910...27M}, we retrieved single-exposure W1 and W2 photometry from the NEOWISE-R catalog for each target, together with reference stars within a 2~deg radius, via the IRSA archive. Recommended quality flags were applied to ensure reliable measurements.

For the variability analysis, two complementary metrics were adopted. The Stetson $J$ index \citep{1996PASP..108..851S} was used to quantify correlated brightness variations between the W1 and W2 bands, while the reduced chi-square, $\chi^2_{\rm red}$, was computed for each band separately to assess whether the observed flux scatter exceeds that expected from photometric uncertainties \citep{2017MNRAS.464..274S}.

The Stetson $J$ index is defined as:
\begin{equation}
    S_J = \frac{\sum_{k=1}^{n} w_k \, \mathrm{sgn}(P_k) \sqrt{|P_k|}}{\sum_{k=1}^{n} w_k}
\end{equation}
where $n$ is the number of epochs with simultaneous observations in both bands, $w_k$ is the weight for the $k$-th epoch (set to 1 in this work), and $P_k$ is the product of normalized residuals in the 2 bands at epoch $k$, defined as:
\begin{equation}
    P_k = \left( \frac{W1_k - \overline{W1}}{\sigma_{W1,k}} \right) \times \left( \frac{W2_k - \overline{W2}}{\sigma_{W2,k}} \right)
\end{equation}
where $W1_k$ and $W2_k$ are the magnitudes in the W1 and W2 bands at epoch $k$, $\sigma_{W1,k}$ and $\sigma_{W2,k}$ are the corresponding photometric uncertainties, and $\overline{W1}$ and $\overline{W2}$ are the mean magnitudes in each band.

The reduced \( \chi^2_{\text{red}} \) is defined as:
\begin{equation}
    \chi^2_{\mathrm{red}} = \frac{1}{N - 1} \sum_{k=1}^{N} \left( \frac{W_k - \overline{W}}{\sigma_k} \right)^2
\end{equation}
where $W_k$ and $\sigma_k$ are the measured magnitude and uncertainty at epoch $k$, and $\overline{W}$ is the weighted mean magnitude over $N$ epochs.

We iteratively performed a 3$\sigma$ clipping on the distributions of the Stetson $J$ index and $\chi^2_{\rm red}$ for each region to derive the mean $D$ and standard deviation $\sigma_D$. The threshold for variability was then defined as $D + 5\sigma_D$. A source is considered variable if any of the indicators (Stetson $J$, $\chi^2_{\rm red, W1}$, or $\chi^2_{\rm red, W2}$) exceeds this threshold.

\begin{figure}[h]
    \centering
    \includegraphics[width=1\linewidth]{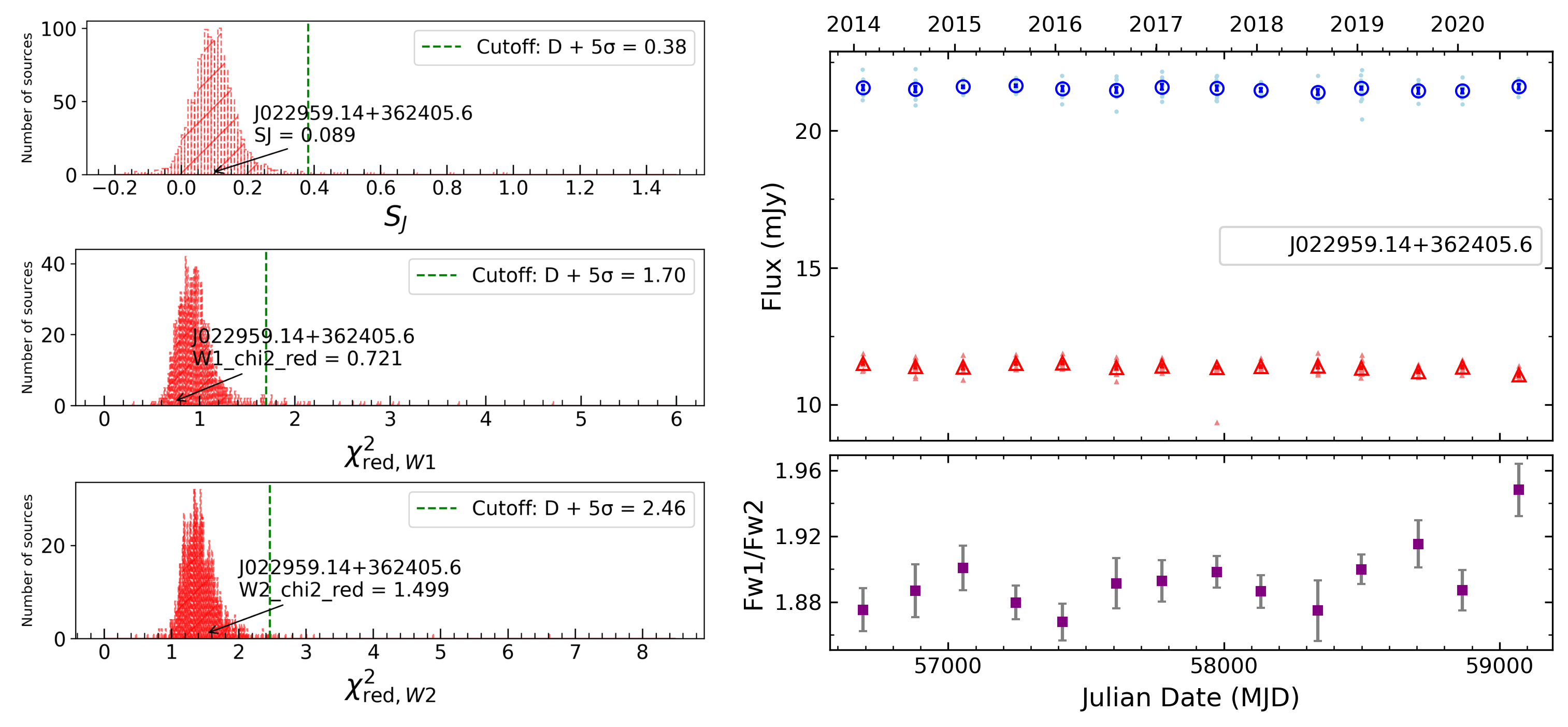}
    \caption{The left panel shows the variability indices and thresholds for J022959.14+362405.6. The green dashed line represents the variability threshold, and the arrow indicates the position of the target source. The right panel shows the seasonal averages of W1 and W2 fluxes for the sample, with blue circles and red triangles representing W1 and W2, respectively. Error bars indicate the uncertainties of the seasonal averages. Purple squares show the W1/W2 flux ratio and its associated uncertainties.}
    \label{fig:light variations}
\end{figure}

The results indicate that none of the sources in Sample~D exhibit significant infrared variability, suggesting that the observed infrared excess is unlikely to be driven by intrinsic stellar variability, and that no evidence for substantial changes in the infrared-emitting region is found, consistent with emission arising from a stable circumstellar dust disc. Figure~\ref{fig:light variations}presents the distributions of the three indices and the long-term flux variations in the W1 and W2 bands for the source J022959.14+362405.6, illustrating the stability of its infrared emission.

\subsubsection{Wide Companions}
Some infrared-excess sources may belong to multiple-star systems, where the presence of companions can reveal the dynamical characteristics of the system. For instance, companions may provide dynamical constraints on the primary star, thereby affecting the orbital evolution of the system and the stability of the debris disc. To investigate whether potential wide-separation companions exist among the sources in Sample~D, we searched the Gaia~DR3 catalogue for objects with similar proper motions and distances to the target stars.

Following ~\citet{2017MNRAS.472..675A}, to ensure the reliability of astrometric solutions, only sources satisfying the following parallax and proper motion quality criteria were considered:

\begin{itemize}
  \item Parallax signal-to-noise ratio: $\pi / \sigma_\pi > 5$
  \item Proper motion signal-to-noise ratio: \( \frac{ \sqrt{ \mu_{\alpha*}^2 + \mu_{\delta}^2 } }{ \sqrt{ \sigma_{\mu_{\alpha*}}^2 + \sigma_{\mu_{\delta}}^2 } } > 5 \).where \( \mu_{\alpha*} \) and \( \mu_\delta \) are the proper motions in right ascension and declination, respectively.
\end{itemize}

The search for co-moving companions was performed following the criteria proposed by ~\citet{2020MNRAS.496.5176D}. This method takes into account differences in astrometric parameters and incorporates an orbital velocity term to assess whether a given pair of objects is likely to form a gravitationally bound system. The criterion is expressed as:

\begin{equation}
n^2_\sigma = \frac{(\mu_{\alpha*,1} - \mu_{\alpha*,2})^2}{\sigma_{\mu_{\alpha*,1}}^2 + \sigma_{\mu_{\alpha*,2}}^2 + \mu^2_{\mathrm{orb},\alpha}} + \frac{(\mu_{\delta,1} - \mu_{\delta,2})^2}{\sigma_{\mu_{\delta,1}}^2 + \sigma_{\mu_{\delta,2}}^2 + \mu^2_{\mathrm{orb},\delta}} + \frac{(\pi_1 - \pi_2)^2}{\sigma_{\pi_1}^2 + \sigma_{\pi_2}^2}
\end{equation}
where $\mu_{\mathrm{orb},\alpha} = \mu_{\mathrm{orb},\delta}$ are the projected orbital motions at 1000\,au, estimated as:
\begin{equation}
\mu_{\mathrm{orb}} = 29.8\,\mathrm{km\,s}^{-1} \times \left( \frac{1000}{4.74\,d_{\mathrm{cluster}}} \right) \sqrt{\frac{1}{3}} \left( \frac{M_1 + M_2}{\rho} \right)^{1/2}
\end{equation}
In this context, the unit of \( \mu_{\mathrm{orb}} \) is mas/yr. The first term in the formula converts orbital velocity from m/s to mas/yr, the second term accounts for the projection of the 3D motion, and the third term represents the total orbital velocity of the binary system. The cluster distance \( d_{\mathrm{cluster}} \) corresponds to the distance of the primary target. 
Finally, a threshold of $n_\sigma < 5$ was adopted to identify candidate co-moving companions. 

The results show that, within Sample~D, only J033941.18+231726.7 and J010859.11+385050.9 each match one possible co-moving companion, with Gaia DR3 source IDs Gaia~DR3~65113529568721152 and Gaia~DR3~370243090202595712, respectively. The projected separations of these pairs are 11\,155.62~au and 28\,247.25~au. Although companions at comparable separations have been reported in previous studies—for example, \citet{2015ApJ...798...86Z} identified a companion to 2M1337 with a projected separation of $\sim$12\,000~au, these separations are significantly larger than the typical range considered capable of maintaining long-term gravitational binding ($\lesssim 6000$~au; \citealt{2021MNRAS.506.2269E}). Therefore, the physical association of these pairs cannot be established with the adopted criteria. They may represent extremely wide binaries, chance alignments, or non-bound members of the same moving group, and further dynamical analysis is required to distinguish between these possibilities.

\subsubsection{Disc Properties}

As shown in Figure~\ref{fig:SED}, the SEDs of the first three sources in Sample~D exhibit infrared excesses in both the W3 and W4 bands; these sources were further analyzed using binary fits with VOSA to derive the dust temperatures and fractional luminosities of their debris discs. 

Debris disks are generally regarded as optically thin dust structures with low fractional luminosities and little residual gas. Both observational and theoretical studies show that they are collisionally produced and remain optically thin at infrared and sub-millimeter wavelengths \citep{2008ARA&A..46..339W,2010RAA....10..383K}. Under this assumption, dust mass, locations, and radiating area can be estimated from radiative equilibrium without significant self-shielding effects \citep{2018ApJ...855...56W,2020A&A...640A..12O}.

\begin{itemize}
\item Dust Mass: To estimate the dust mass surrounding the stars, we adopt a simplified model based on near-infrared fluxes, ignoring scattering and temperature gradients. For each target, the stellar Planck blackbody intensity $B_\lambda(T_{\mathrm{dust}})$ at $\lambda = 2.2031\,\mu\mathrm{m}$ (22031~\AA) is calculated using its temperature $T_{\text{dust}}$, with units of $\mathrm{erg\,s^{-1}\,cm^{-2}\,\AA^{-1}\,sr^{-1}}$. The circumstellar flux $F_\lambda$ and stellar distance $D$ are then used to estimate the dust mass $M_{\mathrm{dust}}$ following \citep{2019ApJ...874..141U}.
\end{itemize}
\begin{equation}
M_{\rm dust} = \frac{F_\lambda D^{2}}
{\kappa_\lambda B_\lambda\!\left(T_{\rm dust}\right)}
\end{equation}

where $\kappa_\lambda$ is the dust absorption coefficient, set to $\kappa_\lambda = 3800~{\rm cm^2~g^{-1}}$ at 2.2~$\mu$m \citep{2011ApJ...728..143S}, $F_\lambda$ is the observed flux, and $D$ is the stellar distance in centimetres.

\begin{itemize}
\item Disc Radius and Radiating Area: The dust radius $R_{\rm disc}$ is estimated assuming the dust is in radiative equilibrium and behaves as a blackbody, following \citet{2008ARA&A..46..339W}:
\end{itemize}

\begin{equation}
R_{\rm disc} = \left( \frac{278~{\rm K}}{T_{\rm dust}} \right)^2 \sqrt{\frac{L_\ast}{L_\odot}}~ .
\end{equation}

where $T_{\rm dust}$ is the dust temperature in K, and $L_\ast$ is the stellar luminosity in units of $L_\odot$.

The total radiating cross-section of the dust, $\sigma_{\rm tot}$, is calculated from the fractional luminosity $f_{\rm d}$ and the disc radius:

\begin{equation}
\sigma_{\rm tot} = 4 \pi R_{\rm disc}^2 f_{\rm d} .
\end{equation}

 Table~\ref{tab:disk} summarises the derived disc parameters including $T_{\rm dust}$, $f_{\rm d}$, $R_{\rm disc}$, and $\sigma_{\rm tot}$ for 3 debris disc candidates.
\begin{table}[ht]
\centering
\caption{Derived disk parameters for 3 debris disk candidates exceed in W3 and W4.}
\begin{tabular}{c c c c c c}
\hline
\multicolumn{1}{c}{\textbf{AllWISE}} & \textbf{Teff} & \textbf{fd} & \textbf{Mass} & \textbf{Radius} & \textbf{$\sigma_{\text{tot}}$} \\
\cline{2-6}  
& K & & M$_{\odot}$ & au & au$^2$\\
\hline
J083717.38+512013.7 & 200 & 1.97E-03 & 1.43E-12 & 1.4026 & 0.0487 \\
J022959.14+362405.6 & 150 & 5.33E-03 & 3.71E-12 & 1.3739 & 0.1264 \\
J033941.18+231726.7 & 200 & 5.88E-04 & 1.34E-12 & 2.5117 & 0.0466 \\
\hline
\end{tabular}
\label{tab:disk}
\end{table}

\section{Summary}
\label{sect:Summary}
We have conducted a systematic search for debris disk candidates around FGK stars within 150 pc by combining a spectroscopically identified stellar sample from LAMOST DR12 with \textit{Gaia} astrometry and multi-band infrared photometry from 2MASS and AllWISE. Main-sequence stars were selected based on spectral type classification from LAMOST and color-based criteria, and infrared excesses were identified through SED fitting and subjected to a multi-layer validation process, which included conservative, source-by-source checks such as visual image inspection, positional offset analysis, close-binary screening, and \textit{Gaia} Figure of Merit checks.

This approach resulted in a final sample of 12 debris disk candidates. A thorough literature search revealed that two of these candidates had been previously confirmed as debris disk systems, while the remaining ten represent new identifications. Based on the work from LAOSMT DR7 \citep{2023A&A...673A.155Q}, suggests that most of the stars are several billion years old. Long-term NEOWISE monitoring in the W1 and W2 bands reveals no significant variability, consistent with a circumstellar origin of the observed infrared excess. A search for co-moving stellar companions using \textit{Gaia} DR3 astrometry identified potential companions for only two candidates, both at very large projected separations ($\gtrsim 10^4$~au). The nature of these pairs, whether they are bound systems or chance alignments, requires further investigation through additional kinematic and astrophysical constraints. Among the 12 candidates, three show significant infrared excess in both the W3 and W4 bands, which allows for estimates of their dust temperatures, characteristic disk radii, and fractional luminosities that are consistent with debris disk systems. 

This work provides a complementary perspective by identifying a small but reliable sample of debris disk candidates around FGK stars. The homogeneous LAMOST spectroscopic data enable future studies of stellar properties and activity, and the identified systems are promising targets for follow-up observations with next-generation facilities such as the James Webb Space Telescope (JWST).

\begin{acknowledgements}
The authors thank the referee for careful reading of the manuscript and feedback, which improved the clarity of this paper. This work was supported in part by the National Natural Science Foundation of China under grant U1631109. This work was also supported in part by the Guizhou Provincial Basic Research Program (Natural Science) (No. MS[2025]694), the Guizhou Provincial Major Scientific and Technological Programs (No. XKBF (2025)011, XKBF (2025)010) and the Guizhou Provincial Science and Technology Projects (No. QKHFQ[2023]003 and QKHPTRC-ZDSYS[2023]003). This work is based on data from the Large Sky Area Multi-Object Fiber Spectroscopic Telescope (LAMOST) DR12, operated by the National Astronomical Observatories, Chinese Academy of Sciences. This publication used data from the ALLWISE and NEOWISE missions, funded by NASA, and based on WISE observations, a collaboration between the University of California, Los Angeles, and the Jet Propulsion Laboratory/California Institute of Technology. Additionally, data from the 2MASS survey, a collaboration between the University of Massachusetts and the Infrared Processing and Analysis Center/California Institute of Technology, were utilized. Photometric and astrometric data from the ESA Gaia mission, processed by the Gaia Data Processing and Analysis Consortium (DPAC), were also used. The Gaia mission website is (\url{https://www.cosmos.esa.int/gaia}). This research also made use of the NASA/IPAC Infrared Science Archive, the VizieR catalogue, SIMBAD database at CDS, Strasbourg, and the VOSA tool under the Spanish Virtual Observatory project (\url{https://svo.cab.inta-csic.es}). The Python ecosystem, including ASTROPY, ASTROQUERY \citep{2013A&A...558A..33A}, PANDAS \citep{pandas2020}, NUMPY, SCIPY \citep{2020Natur.585..357H}, and MATPLOTLIB \citep{2007CSE.....9...90H}, was also used.
\end{acknowledgements}

\appendix
\section{Sources Excluded Based on Visual Inspection}
\label{sec:Sources Excluded Based on Visual Inspection}
This appendix shows the visual inspection results for the 30 sources excluded from Sample~C. Based on AllWISE W1 and W4 images (discuss in Section~\ref{sect:Visual Inspection}), the figures illustrate representative contaminant types and support the robustness of the final debris disk sample.

\begin{figure}[htbp]
    \centering
    \includegraphics[width=1\linewidth]{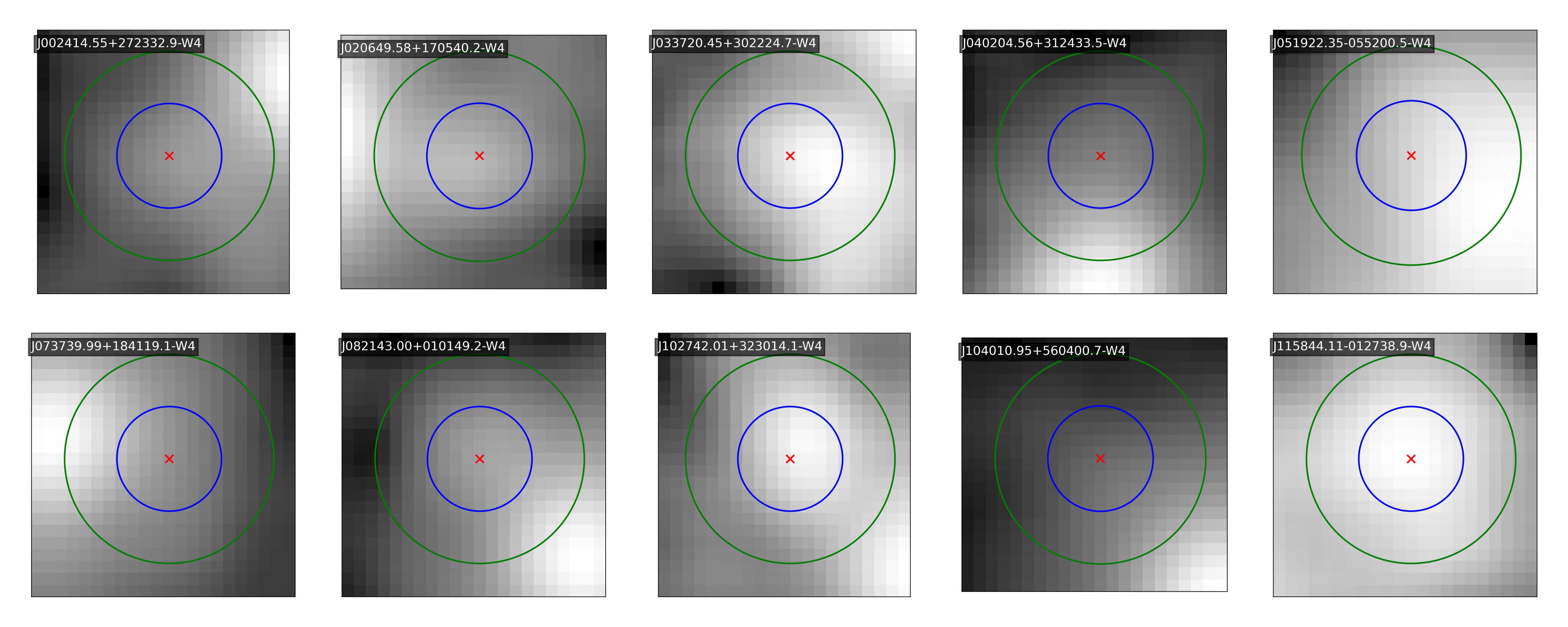}
    \label{fig:placeholder}
\end{figure}

\begin{figure}[htbp]
    \centering
    \includegraphics[width=1\linewidth]{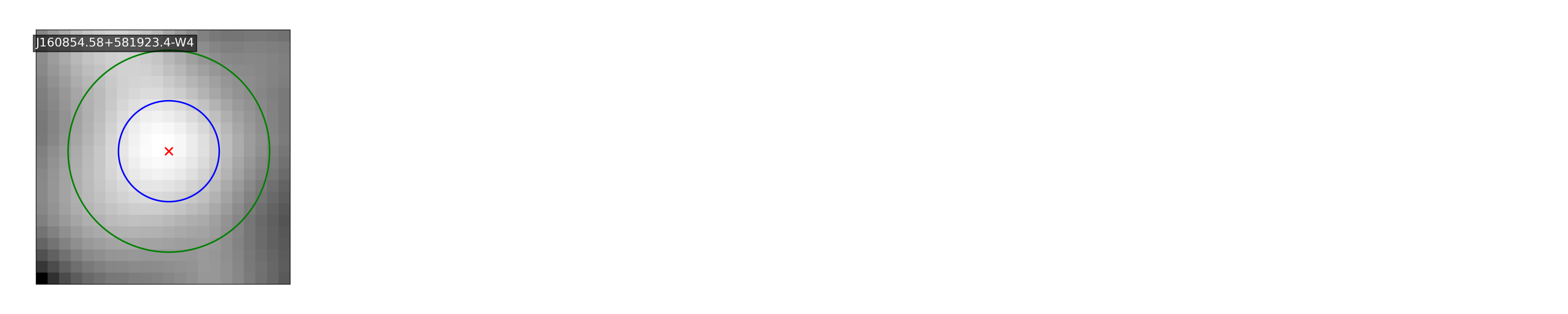}
    \caption{Examples of blended sources identified through visual inspection. The blue and red circles indicate radii of $6^{\prime\prime}$ and $12^{\prime\prime}$ around the target. Nearby infrared sources within $12^{\prime\prime}$ may affect the photometry; thus, these sources are conservatively classified as blended and excluded from the sample.}
    \label{fig:placeholder2}
\end{figure}

\begin{figure}[htbp]
    \centering
    \includegraphics[width=1\linewidth]{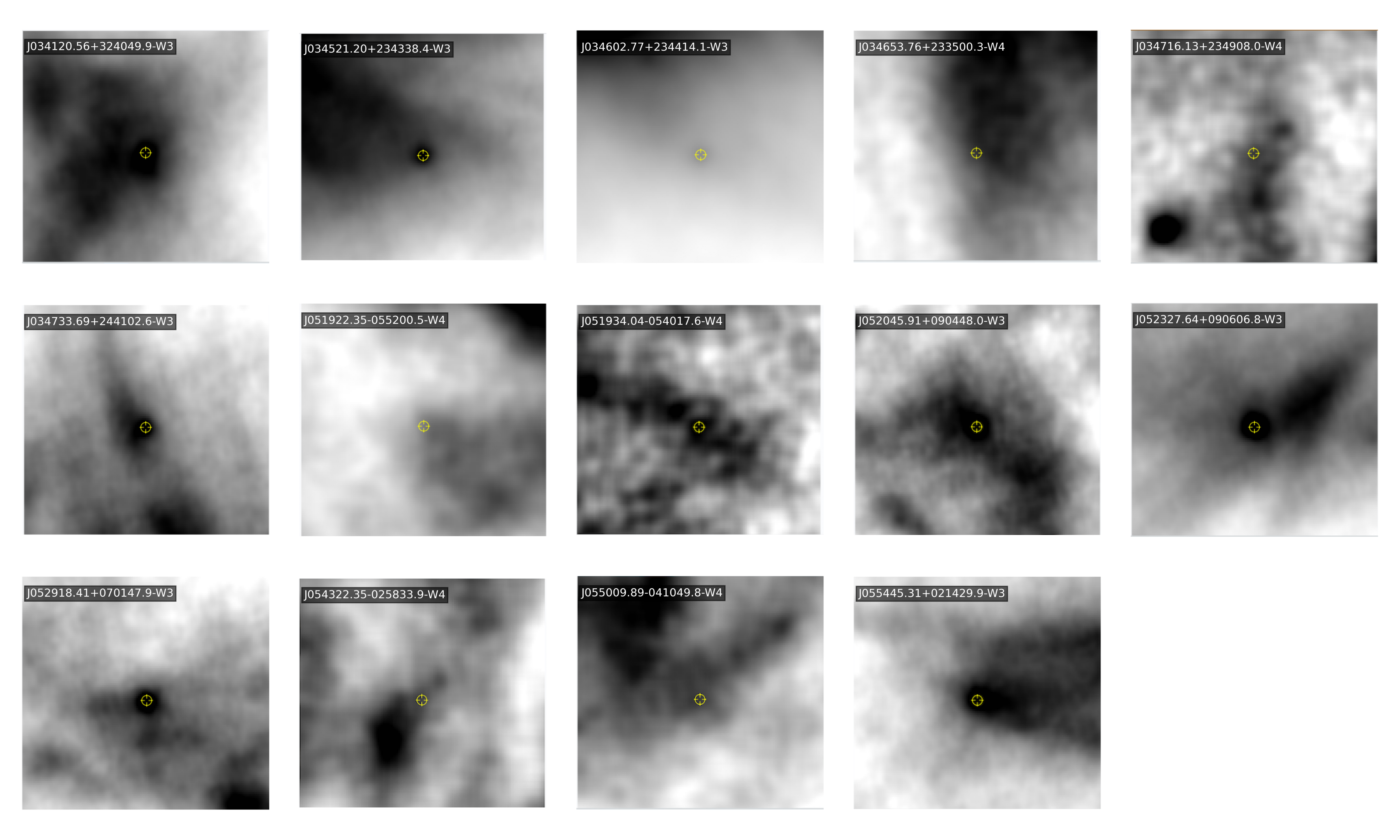}
    \caption{Examples of sources affected by nebular emission in the WISE images, shown in the corresponding infrared-excess bands.}
    \label{fig:Nebular}
\end{figure}

\begin{figure}[H]
    \centering
    \includegraphics[width=1\linewidth]{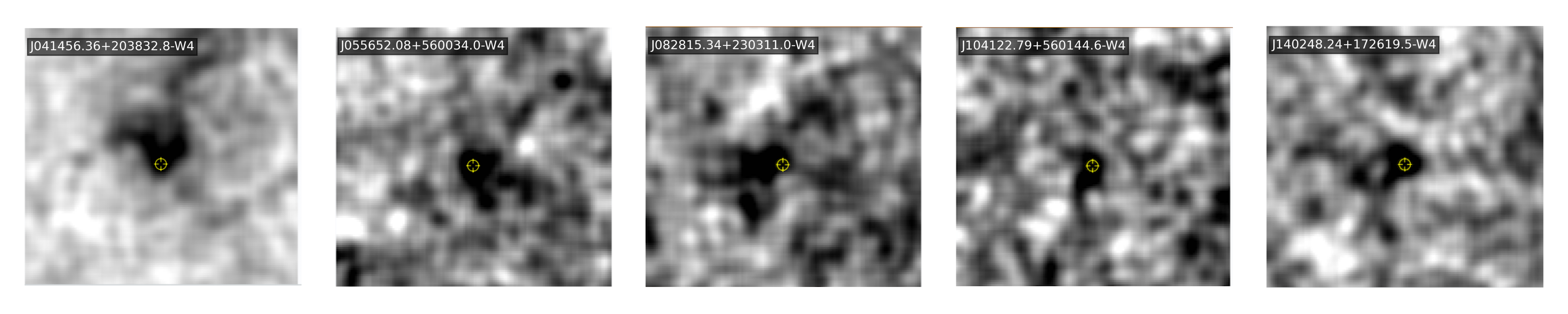}
    \caption{Irregular sources as seen in the corresponding infrared-excess bands.}
    \label{fig:Irregular}
\end{figure}

\bibliography{ms2026-0013}  

\begin{thebibliography}{46}
\providecommand\natexlab[1]{#1}
\providecommand\JournalTitle[1]{#1}

\bibitem[{Andrews} {et~al.}(2017)]{2017MNRAS.472..675A}
{Andrews}, J.~J., {Chanam{\'e}}, J., \& {Ag{\"u}eros}, M.~A. 2017, \mnras, 472,
  675

\bibitem[{Astropy Collaboration} {et~al.}(2013)]{2013A&A...558A..33A}
{Astropy Collaboration}, {Robitaille}, T.~P., {Tollerud}, E.~J., {et~al.} 2013,
  \aap, 558, A33

\bibitem[{Bailer-Jones}(2011)]{2011MNRAS.411..435B}
{Bailer-Jones}, C.~A.~L. 2011, \mnras, 411, 435

\bibitem[{Bayo} {et~al.}(2008)]{2008A&A...492..277B}
{Bayo}, A., {Rodrigo}, C., {Barrado Y Navascu{\'e}s}, D., {et~al.} 2008, \aap,
  492, 277

\bibitem[{Castelli} \& {Kurucz}(2003)]{2003IAUS..210P.A20C}
{Castelli}, F., \& {Kurucz}, R.~L. 2003, in IAU Symposium, Vol. 210, Modelling
  of Stellar Atmospheres, ed. N.~{Piskunov}, W.~W. {Weiss}, \& D.~F. {Gray},
  A20

\bibitem[{Chen} {et~al.}(2014)]{2014ApJS..211...25C}
{Chen}, C.~H., {Mittal}, T., {Kuchner}, M., {et~al.} 2014, \apjs, 211, 25

\bibitem[{Chulkov} \& {Malkov}(2022)]{2022MNRAS.517.2925C}
{Chulkov}, D., \& {Malkov}, O. 2022, \mnras, 517, 2925

\bibitem[{Cotten} \& {Song}(2016)]{2016ApJS..225...15C}
{Cotten}, T.~H., \& {Song}, I. 2016, \apjs, 225, 15

\bibitem[{Cruz-Saenz de Miera} {et~al.}(2014)]{2014MNRAS.437..391C}
{Cruz-Saenz de Miera}, F., {Chavez}, M., {Bertone}, E., \& {Vega}, O. 2014,
  \mnras, 437, 391

\bibitem[{Cui} {et~al.}(2012)]{2012RAA....12.1197C}
{Cui}, X.-Q., {Zhao}, Y.-H., {Chu}, Y.-Q., {et~al.} 2012, Research in Astronomy
  and Astrophysics, 12, 1197

\bibitem[{Da Costa} {et~al.}(2017)]{2017ApJ...837...15D}
{Da Costa}, A.~D., {Canto Martins}, B.~L., {Le{\~a}o}, I.~C., {et~al.} 2017,
  \apj, 837, 15

\bibitem[{Deacon} \& {Kraus}(2020)]{2020MNRAS.496.5176D}
{Deacon}, N.~R., \& {Kraus}, A.~L. 2020, \mnras, 496, 5176

\bibitem[{Deng} {et~al.}(2012)]{2012RAA....12..735D}
{Deng}, L.-C., {Newberg}, H.~J., {Liu}, C., {et~al.} 2012, Research in
  Astronomy and Astrophysics, 12, 735

\bibitem[{Dennihy} {et~al.}(2020)]{2020ApJ...891...97D}
{Dennihy}, E., {Farihi}, J., {Gentile Fusillo}, N.~P., \& {Debes}, J.~H. 2020,
  \apj, 891, 97

\bibitem[{El-Badry} {et~al.}(2021)]{2021MNRAS.506.2269E}
{El-Badry}, K., {Rix}, H.-W., \& {Heintz}, T.~M. 2021, \mnras, 506, 2269

\bibitem[{Harris} {et~al.}(2020)]{2020Natur.585..357H}
{Harris}, C.~R., {Millman}, K.~J., {van der Walt}, S.~J., {et~al.} 2020, \nat,
  585, 357

\bibitem[{Hughes} {et~al.}(2018)]{2018ARA&A..56..541H}
{Hughes}, A.~M., {Duch{\^e}ne}, G., \& {Matthews}, B.~C. 2018, \araa, 56, 541

\bibitem[{Hunter}(2007)]{2007CSE.....9...90H}
{Hunter}, J.~D. 2007, Computing in Science and Engineering, 9, 90

\bibitem[{Jarrett} {et~al.}(2011)]{2011ApJ...735..112J}
{Jarrett}, T.~H., {Cohen}, M., {Masci}, F., {et~al.} 2011, \apj, 735, 112

\bibitem[{Kennedy} \& {Wyatt}(2014)]{2014MNRAS.444.3164K}
{Kennedy}, G.~M., \& {Wyatt}, M.~C. 2014, \mnras, 444, 3164

\bibitem[{Krivov}(2010)]{2010RAA....10..383K}
{Krivov}, A.~V. 2010, Research in Astronomy and Astrophysics, 10, 383

\bibitem[{Lingam} \& {Loeb}(2017)]{2017ApJ...848...41L}
{Lingam}, M., \& {Loeb}, A. 2017, \apj, 848, 41

\bibitem[{Luo} {et~al.}(2015)]{2015RAA....15.1095L}
{Luo}, A.-L., {Zhao}, Y.-H., {Zhao}, G., {et~al.} 2015, Research in Astronomy
  and Astrophysics, 15, 1095

\bibitem[{Mizuki} {et~al.}(2024)]{2024AJ....167..275M}
{Mizuki}, T., {Momose}, M., {Aizawa}, M., \& {Kobayashi}, H. 2024, \aj, 167,
  275

\bibitem[{Mo{\'o}r} {et~al.}(2021)]{2021ApJ...910...27M}
{Mo{\'o}r}, A., {{\'A}brah{\'a}m}, P., {Szab{\'o}}, G., {et~al.} 2021, \apj,
  910, 27

\bibitem[{Olofsson} {et~al.}(2020)]{2020A&A...640A..12O}
{Olofsson}, J., {Milli}, J., {Bayo}, A., {Henning}, T., \& {Engler}, N. 2020,
  \aap, 640, A12

\bibitem[{Patel} {et~al.}(2014)]{2014ApJS..212...10P}
{Patel}, R.~I., {Metchev}, S.~A., \& {Heinze}, A. 2014, \apjs, 212, 10

\bibitem[{Qian} {et~al.}(2019)]{2019RAA....19...64Q}
{Qian}, S.-B., {Shi}, X.-D., {Zhu}, L.-Y., {et~al.} 2019, Research in Astronomy
  and Astrophysics, 19, 064

\bibitem[{Queiroz} {et~al.}(2023)]{2023A&A...673A.155Q}
{Queiroz}, A.~B.~A., {Anders}, F., {Chiappini}, C., {et~al.} 2023, \aap, 673,
  A155

\bibitem[{Rieke} {et~al.}(2005)]{2005ApJ...620.1010R}
{Rieke}, G.~H., {Su}, K.~Y.~L., {Stansberry}, J.~A., {et~al.} 2005, \apj, 620,
  1010

\bibitem[{Shirley} {et~al.}(2011)]{2011ApJ...728..143S}
{Shirley}, Y.~L., {Huard}, T.~L., {Pontoppidan}, K.~M., {et~al.} 2011, \apj,
  728, 143

\bibitem[{Sierchio} {et~al.}(2010)]{2010ApJ...712.1421S}
{Sierchio}, J.~M., {Rieke}, G.~H., {Su}, K.~Y.~L., {et~al.} 2010, \apj, 712,
  1421

\bibitem[{Sokolovsky} {et~al.}(2017)]{2017MNRAS.464..274S}
{Sokolovsky}, K.~V., {Gavras}, P., {Karampelas}, A., {et~al.} 2017, \mnras,
  464, 274

\bibitem[{Stetson}(1996)]{1996PASP..108..851S}
{Stetson}, P.~B. 1996, \pasp, 108, 851

\bibitem[{Su} {et~al.}(2022)]{2022ApJS..261...26S}
{Su}, T., {Zhang}, L.-y., {Long}, L., {et~al.} 2022, \apjs, 261, 26

\bibitem[{Suazo} {et~al.}(2024)]{2024MNRAS.531..695S}
{Suazo}, M., {Zackrisson}, E., {Mahto}, P.~K., {et~al.} 2024, \mnras, 531, 695

\bibitem[Team(2020)]{pandas2020}
Team, P.~D. 2020, pandas, \url{https://doi.org/10.5281/zenodo.3509134}

\bibitem[{Trilling} {et~al.}(2008)]{2008ApJ...674.1086T}
{Trilling}, D.~E., {Bryden}, G., {Beichman}, C.~A., {et~al.} 2008, \apj, 674,
  1086

\bibitem[{Utomo} {et~al.}(2019)]{2019ApJ...874..141U}
{Utomo}, D., {Chiang}, I.-D., {Leroy}, A.~K., {Sandstrom}, K.~M., \&
  {Chastenet}, J. 2019, \apj, 874, 141

\bibitem[{Wang} {et~al.}(2020)]{2020ApJ...891...23W}
{Wang}, R., {Luo}, A.-L., {Chen}, J.-J., {et~al.} 2020, \apj, 891, 23

\bibitem[{Wang} {et~al.}(1996)]{1996ApOpt..35.5155W}
{Wang}, S.-G., {Su}, D.-Q., {Chu}, Y.-Q., {Cui}, X., \& {Wang}, Y.-N. 1996,
  \ao, 35, 5155

\bibitem[{Wilner} {et~al.}(2018)]{2018ApJ...855...56W}
{Wilner}, D.~J., {MacGregor}, M.~A., {Andrews}, S.~M., {et~al.} 2018, \apj,
  855, 56

\bibitem[{Wu} {et~al.}(2011)]{2011RAA....11..924W}
{Wu}, Y., {Luo}, A.-L., {Li}, H.-N., {et~al.} 2011, Research in Astronomy and
  Astrophysics, 11, 924

\bibitem[{Wyatt}(2008)]{2008ARA&A..46..339W}
{Wyatt}, M.~C. 2008, \araa, 46, 339

\bibitem[{Zhao} {et~al.}(2012)]{2012RAA....12..723Z}
{Zhao}, G., {Zhao}, Y.-H., {Chu}, Y.-Q., {Jing}, Y.-P., \& {Deng}, L.-C. 2012,
  Research in Astronomy and Astrophysics, 12, 723

\bibitem[{Zuckerman}(2015)]{2015ApJ...798...86Z}
{Zuckerman}, B. 2015, \apj, 798, 86

\end{thebibliography}

\end{document}